\documentclass[amsmath,amssymb,superscriptaddress,nofootinbib]{revtex4}
\usepackage[dvipdfmx]{graphicx}
\usepackage{dcolumn}% Align table columns on decimal point
\usepackage{bm}% bold math
\usepackage{color}
\usepackage{eucal}
\usepackage{ulem}
\usepackage{color}

\def\beq{\begin{equation}}
\def\eeq{\end{equation}}
\def\beqa{\begin{eqnarray}}
\def\eeqa{\end{eqnarray}}

\def\degC{\kern-.2em\r{}\kern-.3em C}

\begin{document}

\title{Direct Determination of Exciton Wave Function Amplitudes \\
by the Momentum-Resolved Photo-Electron Emission Experiment }

\author{Hiromasa Ohnishi}
\email{hohnishi@tsuruoka-nct.ac.jp}
\affiliation{Department of Creative Engineering, 
National Institute of Technology (NIT), Tsuruoka College, 104 Sawada, Inooka, Tsuruoka, 
Yamagata, 997-8511,Japan}

\author{Norikazu Tomita}
\affiliation{Department of Physics, Yamagata University, 1-4-12 Kojirakawa, Yamagata, 
990-8560, Japan }
  
\author{Keiichiro Nasu}
\affiliation{Institute of Materials Structure Science, High Energy Accelerator Research
Organization (KEK), 1-1 Oho, Tsukuba, 305-0801, Japan}

\begin{abstract}
We study conceptional problems of  a photo-electron emission (PEE) process
from a free exciton in insulating crystals. 
In this PEE process, only the electron constituting the exciton is suddenly emitted out of the crystal, 
while the hole constituting the exciton is still left inside 
and forced to be recoiled back to its original valence band. 
This recoil on the hole is surely reflected in the spectrum of the PEE 
with a statistical distribution along the momentum-energy curve of the valence band. 
This distribution is nothing but the square of the exciton wave function amplitude,  
since it shows how the electron and the hole are originally bound together. 
Thus, the momentum-resolved PEE can directly determine the exciton wave function. 
These problems are clarified, 
taking the $\Gamma$ and the saddle point excitons in GaAs, as typical examples.
New PEE experiments are also suggested.
\end{abstract}

\keywords{momentum-resolved photo-electron emission; exciton wave function.}

\maketitle

\section{Introduction}\label{sec:intro}

The momentum- or the angle-resolved photo-electron emission (PEE) spectroscopy 
is a powerful method to elucidate the energy band dispersion of solids.
Nowadays, by combining this method with the time-resolved tequnique,
we can see the temporal evolution of the energy distribution function in the momentum space,
being able to detect the ultrafast transient dynamics of electrons directly \cite{tanimuraprl}.
Then, the basic experimental tequniqe and theoretical understanding of this method 
have already been established, with a great contibution to the materials science 
\cite{tomita,arpesreview}.
Here, however, we explore a new functionlity of this momentum-resolved PEE
spectroscopy.

In most cases of this PEE, we can tacitly assume that
the initial one-electron energy level, from which the electron starts to be photo-excited and emitted,
is definite, at any rate.
If the starting state is an ordinary exciton, 
freely moving in an insulating crystal, however, such an initial one-electron energy level
is indefinite in principle, even though the exciton itself has a well-defined 
energy level and a momentum.
The free exciton is a bound state between an electron and a hole only in their relative space,
being a superposition of various electron-hole pair excitations,
extending over the wide energy region of the conduction and valence bands \cite{knox} $^{,}$ \cite{toyozawa}.
Thus, the starting one-electron energy level of this PEE is, in principle, indefinite.

By the PEE from this free excition, 
only the electron is suddenly emitted out of the insulating crystal,
while the hole is still left inside, and recoiled back to its original valence band.
This recoil on hole has a statistical distribution along the momentum-energy curve
of the valence band and surely appears in the spectrum of the PEE.
This distribution is nothing but the square of the exciton wave function amplitude 
in the momentum space,
since it shows how the electron and the hole are bound together.
Thus, the momentum-resolved PEE can determine the exciton wave function amplitude, directly.
Incidentally, it is already well known that the PEE can never determine the phase of the emitted electron waves.

At present,
we have no other method to determine the exciton wave function directly
by an experimental observation, although we can theoretically calculate it, by assuming an
effective coulombic attraction between the electron and the hole, as well as their energy band
dispersions \cite{theory}. Thus, this direct observation will contribute to open new aspects of the real electron-hole
correlation in insulating solids, and will provide a novel functionality of the momentum-resolved PEE.

In order to perform this novel experiment, the following quite high level optical
measurement technique is required. At first, a free exciton in an insulating crystal
should be resonantly excited by an intense and temporary short laser pulse. 
A little after this laser pulse excitation having been completed, 
another intense and temporary short laser pulse should be shone 
to make the electron (constituting this exciton) photo-emitted out of the crystal. 
Finally, the energy and the momentum of this
photo-emitted electron should be determined quite precisely. 
Thus, the present problem inevitably becomes a quite novel momentum- and time-resolved PEE.

Very recently, such an experiment  
has been performed by Tanimura \textit{et al.} in GaAs \cite{tanimura}. 
In this PEE, in addition to the ordinary $\Gamma$ exciton in the band gap \cite{sturge},
the saddle point exciton just below the L valley has also been quite clearly detected.
Only by the ordinary optical measurements,
the existence of the saddle point exciton in GaAs has been suggested in Refs. \cite{phillips} and \cite{cardona}.
Since this saddle point exciton is the bound state immersed
in the continuum of the electron-hole pair excitation,
it is known to be the Fano resonance state \cite{fano}.

In this paper, 
we study the recoil effect of the hole
in the PEE from the free exciton.
Our goal is to conceptionally show that the momentum-resolved PEE spectrum 
directly reflects the exciton wave function.
The momentum-resolved PEE can evaluate 
the real electron-hole correlation in the exciton,
basically without any theoretical assistance.

For the explanation of this recoil effect, we need to evaluate the exciton wave function.
Here we do it only by a simple method, since it may be enough 
to make our limited and conceptional purpose
of the present paper clearly understood by readers.
Then, prior to the argument of this recoil effect, 
we introduce a phenomenological model for exciton states in GaAs.
We reproduce the known experimental results as simple as possible.
%since our purpose is not to make any theoretical progress in the exciton research.
We will be concerned only with resonant excitations to the $\Gamma$ exciton or the 
saddle point one.
Even in the cases of various non-resonant excitations,
a conventional two-color two-photon PEE will also occur.
However, their intensities are very weak as compared with those of resonant cases,
and hence the non-resonant effects would not affect the essential aspect of the present argument.
 
This paper is organized as follows:
In Sec. \ref{sec:theory}, we introduce a phenomenological model for exciton states in GaAs.
In Sec. \ref{sec:recoil}, we study the recoil effect of the hole in PEE from excitons,
and show the momentum-resolved PEE can evaluate the exciton wave function amplitude directly.
Finally, in Sec. \ref{sec:summary}, we give a summary.

\section{A phenomenological model for exciton in GaAs}\label{sec:theory}
To evaluate the exciton wave functions for both the $\Gamma$ exciton and the saddle point one in GaAs,
we introduce a phenomenological model, wherein
model parameters are adjusted so as to reproduce the experimental results of the
excitons in GaAs.
For this purpose,
we start from the following Hamiltonian $(\equiv H)$,  
\beq
H=H_0 +H_{\rm{eh}},
\label{hdef}
\eeq 
\beq
H_0 \equiv\sum_{\bm k} E_{\rm e} ({\bm k}) 
a^{\dagger}_{\bm k} a_{\bm k}+
\sum_{\bm k} E_{\rm h} ({\bm k}) 
b^{\dagger}_{\bm k} b_{\bm k},
\label{h0def}
\eeq
\beq
H_{\rm{eh}}\equiv -\sum_{{\bm l}, \Delta{\bm l}} U_{\rm{eff}} (\Delta {\bm l})
n_{{\rm e},{\bm l}} n_{{\rm h},{\bm l}+\Delta{\bm l}},
\label{hehdef}
\eeq
where $a^{(\dagger)}_{\bm k}$ and $b^{(\dagger)}_{\bm k}$ are
annihilation (creation) operators for an electron and a hole with wave vector ${\bm k}$, respectively.
Spin index is neglected as we focus on a singlet exciton.
The electron number and the hole one at a lattice site ${\bm l}$ are defined as
$n_{{\rm e},{\bm l}}\equiv a^{\dagger}_{\bm l} a_{\bm l}$ and 
$n_{{\rm h},{\bm l}}\equiv b^{\dagger}_{\bm l} b_{\bm l}$, respectively,
where
$a_{\bm l}(b_{\bm l})=N^{-1/2}\sum_{\bm k} e^{i {\bm k}\cdot{\bm l}} a_{\bm k} (b_{\bm k})$.
$N$ is the total number of the lattice sites.
$E_{\rm e} ({\bm k})$ and $E_{\rm h} ({\bm k})$ are the band dispersions for the conduction electron
and for the valence hole, respectively.
This energy is parametrized into an analytic function,
whose details are given in Appendix A, by referring the result
in Ref. \cite{cardona}.
The modeled band energy along the $\Gamma$-L line is shown in Fig. 1.
In our model, all the energies are referenced from the valence band maximum.
$-U_{\rm eff}(\Delta {\bm l})$ is the effective attraction between an electron at a site ${\bm l}$
and a hole at a site ${\bm l}+\Delta{\bm l}$,
and is an adjustable parameter to be determined so as to reproduce 
the experimentally observed energies of excitons.
The unit of length is the spatial periodicity of the crystal along the $\Gamma$-L direction.

%----------------------------------------------------------------------
%------ Figure ----------------------------------------------------------
\begin{figure}[tbp]
\centering
\includegraphics[width=5cm]{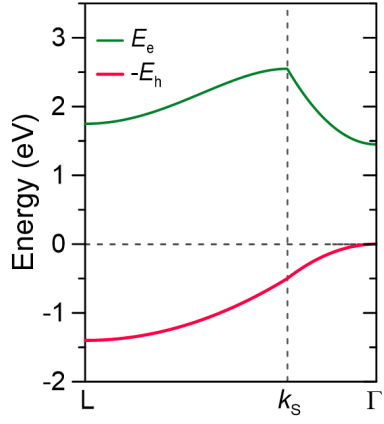} 
\caption{(Color online) Modeled band dispersion for GaAs is shown.
$k_{\rm S}$ represents wave number at the saddle point between the $\Gamma$ valley and the L one.
For more details, see the Appendix.
}
\label{fig1}
\end{figure}
%----------------------------------------------------------------------
%----------------------------------------------------------------------

The excitation spectrum ($\equiv S(E)$) by a light
with the photon energy ($\equiv E$) is given as
\beq
S(E)=\sum_j \mid \langle j \mid P \mid 0 \rangle |^2 \delta (E-E_j),
\label{spectrum}
\eeq
where
$P$ is the polarization or photo-excitation operator, and is given as
\beq
P=N^{-\frac{1}{2}} \sum_{\bm l} (a^{\dagger}_{\bm l} b^{\dagger}_{\bm l}+{\rm h.c.})
=N^{-\frac{1}{2}} \sum_{\bm k} (a^{\dagger}_{\bm k} b^{\dagger}_{-{\bm k}}+{\rm h.c.}).
\label{polarization}
\eeq
$\mid 0 \rangle$ represents the electron-hole vacuum and $\mid j \rangle$ represents
the exact excited state with the energy $E_j$, and hence 
\beq
H \mid j \rangle = E_j \mid j\rangle.
\eeq 
By using the relation
\beq
\delta(E-E_j)=-\frac{1}{\pi} {\rm Im} \left(  
\frac{1}{(E+i\varepsilon)-E_j}
\right),
\eeq
(where $\varepsilon$ is a positive infinitesimal,)
Eq. (\ref{spectrum}) is written as
\beq
S(E)=-\frac{1}{\pi} {\rm Im} \langle 0 \mid {\bm P} 
\frac{1}{(E+i\varepsilon)-{\bm H}} {\bm P} \mid 0 \rangle.
\label{spec1}
\eeq
In the above equation, the right hand side can be expanded with respect to ${\bm H_{\rm eh}}$ as
\begin{equation}
\frac{1}{(E+i\varepsilon)-{\bm H}}   %= \frac{1}{(E+i\varepsilon)-{\bm H}_0 -{\bm H}_{\rm eh}} \\
=\frac{1}{(E+i\varepsilon)-{\bm H}_0} % \\
+\frac{1}{(E+i\varepsilon)-{\bm H}_0} {\bm H}_{\rm eh}\frac{1}{(E+i\varepsilon)-{\bm H}_0} 
+ \cdots.
\end{equation}
This is the Green's function of an excitonic state with an energy $E$, which is given as
\beq
{\bm G} (E+i\varepsilon) \equiv \frac{1}{(E+i\varepsilon)-{\bm H}}
=\frac{1}{(E+i\varepsilon)-{\bm H}_0-{\bm H}_{\rm eh}}.
\label{fullg}
\eeq
This can be also expanded with respect to ${\bm H}_{\rm eh}$ as
\beq
{\bm G}={\bm G}_0 +{\bm G}_0 {\bm H}_{\rm eh} {\bm G}_0
+{\bm G}_0 {\bm H}_{\rm eh} {\bm G}_0 {\bm H}_{\rm eh} {\bm G}_0 + \cdots,
\label{gexpansion}
\eeq
where ${\bm G}_0$ is the Green's function for the non-interacting system, and is 
given as
\beq
{\bm G}_0 (E+i\varepsilon) \equiv \frac{1}{(E+i\varepsilon)-{\bm H}_0}.
\label{g0}
\eeq
%Thus, the present method is corresponding to the ladder approximation. 
By using this Green's function,
the excitation spectrum, Eq.(\ref{spec1}), is expressed as
\beq
S(E)=- \frac{1}{\pi} {\rm Im} \langle 0 \mid {\bm P} {\bm G} (E+i\varepsilon) {\bm P} \mid 0 \rangle.
\label{spectrumg}
\eeq
If we take only the zeroth order term of ${\bm G}$,
$S(E)$ becomes the combined excitation density of states (DOS) of
the non-interacting electron-hole pair ($\equiv D(E)$) as
\beq
D(E)=- \frac{1}{\pi} {\rm Im} \langle 0 \mid {\bm P} {\bm G}_0 (E+i\varepsilon) {\bm P} \mid 0 \rangle.
\eeq
Then, as a scalar, we can define
\beq
G_0 (E+i\varepsilon) \equiv \langle 0 \mid {\bm P} {\bm G}_0 (E+i\varepsilon){\bm P} \mid 0 \rangle.
\eeq
In the present model, it is explicitly written as
\beq
G_0 (E+i\varepsilon)=\frac{1}{N} \sum_{\bm k} 
\frac{1}{E+i\varepsilon-E_{\rm e} ({\bm k})-E_{\rm h} (-{\bm k})}, 
\eeq
and 
\beq
D(E)=\frac{1}{N} \sum_{\bm k} \delta \left(E-E_{\rm e}({\bm k})-E_{\rm h} (-{\bm k}) \right),
\label{ddef}
\eeq 
with the condition $\int D(E) dE=1$.
We can also define 
\beq
R(E)-i\pi D(E) \equiv G_0 (E+i\varepsilon),
\eeq
where the real part $R(E)$ is connected with $D(E)$ by the the Kramers-Kronig relation as,
\beq
R(E)=\int dE' \frac{D(E')}{E-E'}.
\label{kktrans}
\eeq

To determine the excitonic state so as to reproduce the experimental result,
we consider only the effective on-site coulombic interaction between an electron and a hole,
$U_{\rm{eff}} (\Delta {\bm l}) \to U_{\rm{m}} =U_{\rm{eff}} (0)$.
In order to take original long range nature of the coulombic interaction into account,
we assume this $U_{\rm{m}}$ depends on the resultant exciton.
Hence, we assign $\rm{m}=\Gamma$ for the $\Gamma$ exciton and $\rm{m}=S$ for the saddle point one, respectively.
Then, the Green's function for exciton is now represented as
\beq
G(E)\equiv \langle 0 \mid {\bm P} {\bm G}(E) {\bm P} \mid 0 \rangle 
=\frac{G_0(E)}{1+G_0(E)U_{\rm{m}}}.
\label{gfinal}
\eeq 

%----------------------------------------------------------------------
%------ Figure ----------------------------------------------------------
\begin{figure}[tbp]
\centering
\includegraphics[width=6cm]{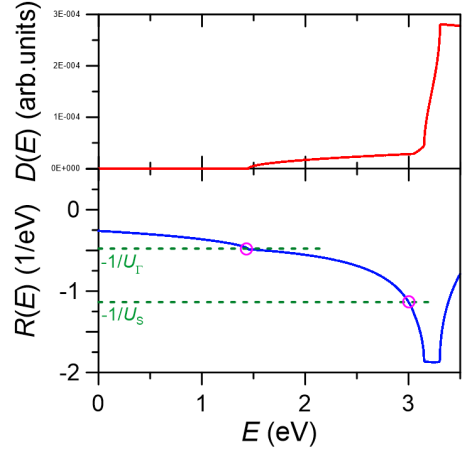} 
\caption{(Color online)The Combined DOS $D(E)$ and $R(E)$ are shown.
The dashed lines are represent $-U^{-1}_{\Gamma}$ and $-U^{-1}_{\rm S}$, which are determined to 
so as to reproduce the experimental energy of excitons.
For more details, see the text.
}
\label{fig2}
\end{figure}
%----------------------------------------------------------------------
%----------------------------------------------------------------------

The estimated $D(E)$ and $R(E)$ are shown in Fig. \ref{fig2},
wherein each exciton state corresponds to the crossing point of the $R(E)$ curve and 
$-U^{-1}_{\rm{m}}$(, $1+R(E)U_{\rm m}=0$). 
To determine the adequate $U_{\rm{m}}$'s,
we use the energy of excitons obtained by the experiment.:
(1) The $\Gamma$ exciton is just 5 meV below the conduction band bottom,
and the band gap energy is set at 1.45 eV in our model.
(2) For the saddle point exciton, its excitation energy is around 3 eV from Ref. \cite{tanimura}.
From these conditions,  we obtain 
$U_{\Gamma}=2.1$ eV for the $\Gamma$ point exciton, and $U_{\rm S}=0.88$ eV for saddle point one,
respectively. 

%Generally speaking,
%the exciton state is formed as a result of competition between kinetic energy increase of the electron-hole pair and 
%energy gain by the attractive interaction between electron and hole.
%In the case of Fano type saddle point exciton,
%an imperfect but good nesting between the conduction band and the valence one along the $\Gamma$-L direction
%is seen in Fig. \ref{fig1}.
%In fact, they are almost parallel with each other.
%It makes the kinetic energy increase smaller than that for the $\Gamma$ exciton.
%It would be important for the stabilization of the saddle point exciton, as suggested in Ref. \cite{cardona}.
%However, we should keep in mind that the present exciton is a three-dimensional state,
%and the anisotropic nature in the three-dimensional Brillouin zone must be included to explain the experimental 
%excitation energy precisely.

In our model, anti-resonance states corresponding to 
the $\Gamma$ exciton and the saddle point one may also exist.
However, they are not relevant to the present purpose.

\begin{figure}[tbp]
\centering
\includegraphics[width=7cm]{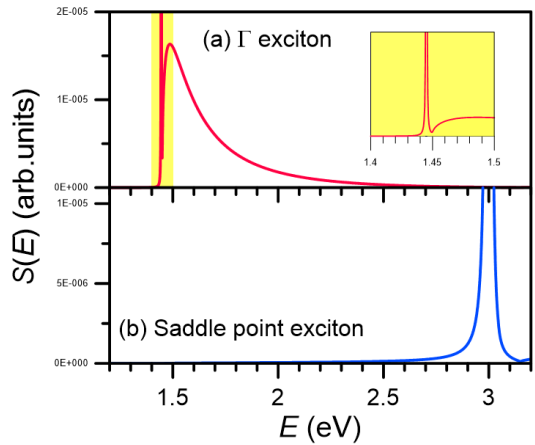} 
\caption{(Color online) The excitation spectrums for (a) $\Gamma$ exciton and (b) Saddle point exciton are shown.
}
\label{fig3}
\end{figure}

The evaluated $S(E)$ is shown in Fig. \ref{fig3}.
Both the $\Gamma$ exciton and the saddle point one basically give Lorentz shapes.
However, Fano type anti-symmetric tails in each low intensity regions
can also be seen clearly.
Since the saddle point exciton is a transient state,
the obtained spectral width reflects the decay rate ($\equiv \Gamma_{\rm S}$).
One can evaluate this $\Gamma_{\rm S}$, 
by assuming the Lorentz form of spectral shape as
\beq
S(E) \propto \frac{1}{(E_{{\rm x},{\rm S}}-E)^2+\Gamma^2_{\rm S}},
\eeq
where $E_{{\rm x},{\rm S}}$ is the energy of the saddle point exciton and 
is set at 3.0 eV in the present model.
By the curve fitting, we can estimate that $\Gamma_{S} \sim 0.1$ eV.
Thus, the present model well reproduces the basic characteristics of the experiments \cite{tanimura, cardona}.

%----------------------------------------------------------------------
%------ Figure ----------------------------------------------------------
%----------------------------------------------------------------------
%----------------------------------------------------------------------

\section{PEE from excitons}\label{sec:recoil}
In this section, we discuss the PEE from the excitons introduced in the preceding section.
Since we focus on the resonant excitation to each exciton in the two-photon process as mentioned before,
we start from the condition that the $\Gamma$ exciton or the saddle point one has already been created.
When this exciton is shone by the light with an energy ($\equiv E_p$),
its electron is excited up to the high energy continuum state within the crystal.
This high energy continuum is the literal ``high energy continuum", in the sense that
the electron with any crystal momentum can find corresponding state at ``any high energy",
and can go out of the crystal, whereafter it is sorted according to the initial crystal momentum and the energy.
This is nothing but the function of the momentum-resolved PEE apparatus.

In the PEE from the exciton,
only the electron is emitted out from the crystal and the hole is still left inside, as mentioned before.
Since the electron and the hole are bound, 
the energy of the emitted electron is shifted by recoil effect of the hole.
Then,  
the energy of emitted electron ($\equiv E_{\rm em} ({\bm k})$), sorted in the crystal momentum ${\bm k}$,
is given from the energy conservation low as
\beq
E_{{\rm x}, m}+E_p=E_{\rm em} ({\bm k})+W+E_{r} ({\bm k}),
\label{eem}
\eeq 
where $W$ is the energy to remove an electron from the crystal, and 
$E_{{\rm x}, m}$ is the energy of exciton with $m=\Gamma$ for the $\Gamma$ exciton and
$m={\rm S}$ for the saddle point one, as mentioned before.
$E_{r} ({\bm k}) (=E_{\rm h}(-{\bm k})>0)$ is the recoil energy of the hole.
$E_{\rm em} ({\bm k})$ depends on $E_{{\rm x},{\rm m}}$, $E_p$ and $W$, as well as $E_{r} ({\bm k})$.
However, these former three quantities can be fixed, once we fix a target exciton state.
Then, by observing $E_{\rm em} ({\bm k})$ experimentally, we can determine $E_{r} ({\bm k})$.
Since the hole is recoiled back to its original valence band,
this recoil on hole has a statistical distribution along the momentum-energy curve
of the valence band. 
The intensity of this statistical distribution corresponds to
the existing probability of the electron-hole pair constituting the exciton.
Thus, the statistical distribution obtained by the recoil on hole is nothing but the square of the 
exciton wave function
amplitude in the momentum space,
since it shows how the electron and hole are bound together.

To see this, we calculate the wave function for the exciton.
The Green's function itself in Eq. (\ref{fullg}) is already a kind of wave function \cite{koster}.
However, here we derive it more straightly.
The eigenstate for the $\Gamma$ exciton with the total momentum zero
can be written as
\beq
\mid \Gamma \rangle =N^{-\frac{1}{2}} \sum_{\bm k} f_{\Gamma} ({\bm k})
a^{\dagger}_{\bm k} b^{\dagger}_{-{\bm k}} \mid 0 \rangle,
\label{eigg}
\eeq
where $ f_{\Gamma} ({\bm k})$ is the wave function of the  $\Gamma$ exciton,
and we determined it variationally.
The relation between the present ${\bm k}$ space representation for the exciton
and the real space one is shown in the Appendix B.

Let us define the Lagrangian $(\equiv {\mathcal L})$ as
\beq
{\mathcal L} (E_{{\rm x},{\Gamma}}) =\langle \Gamma \mid (H_0+H_{{\rm eh}}-E_{{\rm x},\Gamma}) \mid \Gamma \rangle,
\label{vareg}
\eeq
which satisfies
\beq
\frac{\partial {\mathcal L} (E_{{\rm x},{\Gamma}})}{\partial f^{\ast}_{\Gamma}({\bm k})}=0.
\label{mincond1}
\eeq
Since $U_{\rm{eff}} (\Delta {\bm l})$ in Eq. (\ref{hehdef}) is now $U_{\Gamma}$,
from Eqs. (\ref{vareg}) and (\ref{mincond1}), we obtain
\beq
f_{\Gamma} ({\bm k}) \propto \frac{1}{E_{{\rm x},\Gamma}-E_{\rm e}({\bm k})-E_{\rm h} (-{\bm k}) }.
\eeq

In case of the Fano type saddle point exciton, 
due to its transient nature, 
the above variational method cannot be used.
Here, however, we use a simple extrapolation method to calculate the wave function for 
the saddle point exciton.
At first, we tentatively or formally assume that the saddle point exciton has an infinite life time,
$\varepsilon^{-1}$.
In this case, this exciton becomes a good eigenstate of the exciton Hamiltonian, Eq. (\ref{hdef}),
being locally stable(, or an extremal) state, although it is only around the saddle point region
with the energy $\sim E_{{\rm x},{\rm S}}+i\varepsilon$. The eigenstate for the saddle point exciton
with the total momentum zero can be written as
\beq
\mid {\rm S} \rangle =N^{-\frac{1}{2}} \sum_{\bm k} f_{\rm S} ({\bm k})
a^{\dagger}_{\bm k} b^{\dagger}_{-{\bm k}} \mid 0 \rangle,
\label{eigg2nd}
\eeq
where $ f_{\rm S} ({\bm k})$ is the wave function of the saddle point exciton.
Hence, we can determine the wave function $ f_{\rm S} ({\bm k})$ from the locally stable or the extremal condition of
the Lagrangian ${\mathcal L}(E_{{\rm x},{\rm S}})$, which is given by 
\beq
{\mathcal L}(E_{{\rm x},{\rm S}}) =\langle {\rm S} \mid 
\left\{H_0+H_{{\rm eh}}-(E_{{\rm x},{\rm S}}+i\varepsilon)\right\} \mid  {\rm S} \rangle.
\label{varegsad}
\eeq
The locally stable or the extremal condition becomes as
\beq
\frac{\partial {\mathcal L}(E_{{\rm x},{\rm S}})}{\partial f^{\ast}_{\rm S}({\bm k})}=0.
\label{mincond2}
\eeq
By replacing $U_{\rm{S}} (\Delta {\bm l})$ in Eq. (\ref{hehdef}) by $U_{\rm S}$,
we obtain
\beq
f_{\rm S} ({\bm k}) \propto \frac{1}{E_{{\rm x},{\rm S}}+i\varepsilon-E_{\rm e}({\bm k})-E_{\rm h} (-{\bm k}) }.
\eeq
from Eqs. (\ref{varegsad}) and (\ref{mincond2}). 

In the next, only within this context, we introduce its finite life time,
by the extrapolation, $\varepsilon \rightarrow \Gamma_{\rm S}$, 
from the previous Green's function calculation result, and we finally get
\beq
f_{\rm S} ({\bm k}) \propto \frac{1}{E_{{\rm x},{\rm S}}
+i\Gamma_{\rm S}-E_{\rm e}({\bm k})-E_{\rm h} (-{\bm k})}.
\eeq

\begin{figure}[tbp]
\centering
\includegraphics[width=7cm]{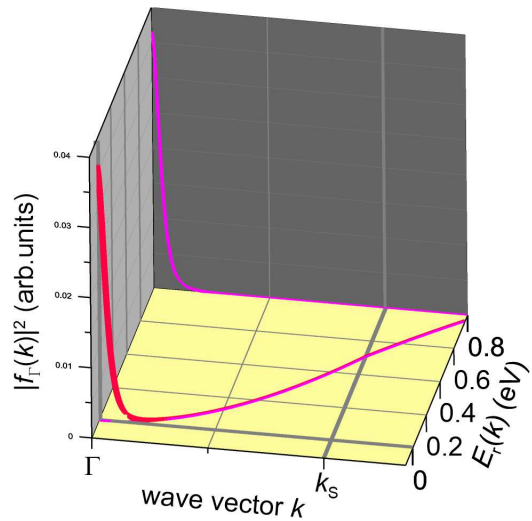} 
\caption{(Color online) Existing probabilities of the $\Gamma$ exciton in momentum space,
$|f_{\Gamma}(k)|^2$, and 
the associated recoil energy of the hole, $E_{r} ({\bm k})$, are shown.
}
\label{fig4}
\end{figure}

\begin{figure}[tbp]
\centering
\includegraphics[width=7cm]{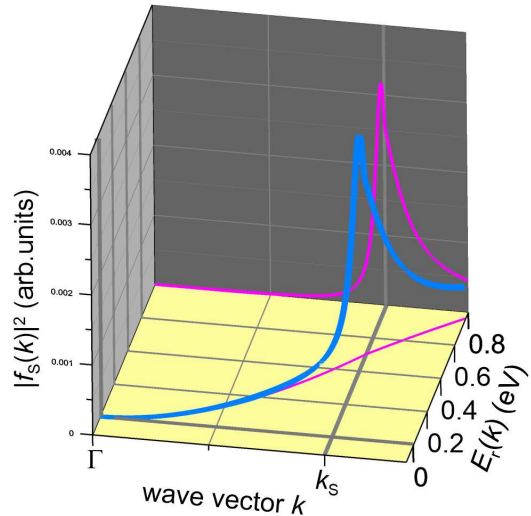} 
\caption{(Color online) Existing probabilities for the saddle point exciton in momentum space, 
$|f_{\rm S}(k)|^2$, and 
the associated recoil energy of the hole, $E_{r} ({\bm k})$, are shown.
}
\label{fig5}
\end{figure}

The calculated $|f_{\rm m}(k)|^2$'s are shown in Figs. \ref{fig4} and \ref{fig5}.
One can see the one-to-one correspondence between 
$|f_{\rm m}(k)|^2$ and $E_{r}(k)$.
For example, the peak of the $|f_{\rm S}(k)|^2$ in Fig. \ref{fig5}
is around the saddle point between the $\Gamma$ valley and the L one,
with $E_{r}(k)\sim0.5$ eV.
Thus, the measured $E_{\rm em} ({\bm k})$ in the momentum-resolved PEE
reflects the wave function of the exciton through $E_{r}({\bm k})$.

The $\Gamma$ exciton is the ordinary one in the energy gap, and with a large electron-hole radius,
or only in the small $k$ region around the $\Gamma$ point.
On the other hand, the saddle point exciton arises mainly from the van Hove singularity of $D(E)$
due to the L valley structure. Because of the electron-hole attraction, however, it has shifted to 
a little below this singular point. Its radius is smaller than that of the $\Gamma$ exciton,
or it is in the larger $k$ region around the $k_{\rm S}$ point. 
It is resonating with the high energy region of the L valley as seen from $D(E)$ in Fig. \ref{fig2},
and becomes the Fano state.

%----------------------------------------------------------------------
%------ Figure ----------------------------------------------------------
%----------------------------------------------------------------------
%----------------------------------------------------------------------
%---------------------------------------------------0-------------------
%------ Figure ----------------------------------------------------------
%----------------------------------------------------------------------
%----------------------------------------------------------------------

Here, we would like to suggest new PEE experiments not only for the excitons in GaAs, 
but also other ones in other materials. 
Especially, we would like to recommend the excitons with a large binding energy 
and (or) a short electron-hole radius. 
In this case, the wave function $f({\bm k})$ extends to a wide region in the Brillouin zone,
as shown in Fig. \ref{fig5}.
Hence, the energy-momentum resolution of the PEE experiment becomes rather easy. 
The direct experimental measurement of this $f({\bm k})$ in the large ${\bm k}$ region is very important, 
since it includes the true information of the real electron-hole attraction at a short distance of each material. 
Although various studies have already been devoted for this problem, 
true natures of electron-hole attraction at a short distance in each material are, 
even at present, still not so well clarified, 
since they are directly related to complicated ``local'' atomic, 
electronic and dielectric characteristics of each material, 
which is not in the ground state, but under a resonant optical ``excitation''. 
By the experiments suggested here, research for this problem is expected to develop greatly.

\section{Summary}\label{sec:summary}
In GaAs, 
we have, thus, shown that the PEE from the exciton can determine 
the square of the exciton wave function amplitude in the momentum space directly, 
reflecting the recoil effect of the hole left in valence band.
Since the PEE spectrum itself is the exciton wave function,
we can obtain the information of the real electron-hole correlation in insulating solids,
basically without any theoretical assistance.
When we look at this thing from an opposite standing point,
the momentum-resolved PEE spectrum from the exciton may provide 
an opportunity to examine a validity of various theoretical and computational methods.
These new aspects are due to the two-body nature of exciton, with the apparent break down
of the one-body picture, in the initial state of the PEE.
On this point, the PEE from the exciton is quite unique, and gives us an interesting insight 
into the many-body effect in quantum systems.

\begin{acknowledgements}
The authors would like to thank K. Tanimura and H. Tanimura for
presenting their research result prior to publication.
This work was partly supported by JSPS Grant-in-Aid 
for Specially Promoted Research, Grant Number 24000006.
\end{acknowledgements}

\appendix

\section{Modeled band dispersion for GaAs}\label{sec:band}

The band dispersion for GaAs is modeled without losing their
basic characteristics, by referring the result
in Ref. \cite{cardona}.
In our parametrization, the unit of length is the lattice constant 
along [111] direction of the GaAs crystal, and
all the energies are referenced from the top of the valence band.
In the reciprocal lattice space, the wave vector along the $\Gamma$-L direction
is represented as $k_{\rm L}$. Two wave vectors normal to $k_{\rm L}$ are
represented as $k_{\rm t1}$ and $k_{\rm t2}$, respectively.
All the energies are given in eV.
In our model, spin and orbital degeneracies are neglected,
since there is no mixing among these degenerate states.

In conduction band, the wave vector at the saddle point between $\Gamma$ valley and L one
is set to $k_{\rm S}\equiv \alpha_{\rm S} \pi$, 
where $k_{\rm L} =0$ for $\Gamma$ point and $\mid k_{\rm L}\mid =\pi$ for L one. 
The band dispersion for $\Gamma$ valley ($0 \le k_{\rm L}, k_{\rm t1},k_{\rm t2} \le k_{\rm S}$) is parametrized with
the following forms:
\beq
E_{\rm e} ({\bm k})=E_{\rm g}+c_g \left\{
\left(\frac{k_{\rm L}}{k_{\rm S}} \right)^2 + 
\left(\frac{k_{\rm t1}}{k_{\rm S}} \right)^2+
\left(\frac{k_{\rm t2}}{k_{\rm S}} \right)^2
\right\},
\label{bdg}
\eeq
where $E_g$ is the band gap.The band dispersion for L valley
($k_{\rm S} \le k_{\rm L} \le \pi$, $0 \le k_{\rm t1},k_{\rm t2} \le k_{\rm S}$)
 is parametrized as
\beq
E_{\rm e} ({\bm k})=E_{g2}+c_{L1}
\left\{ 
d_L \left( \frac{\pi-k_{\rm L}}{\alpha_L \pi}\right)^2 -\left( \frac{\pi-k_{\rm L}}{\alpha_L \pi} \right)^4
\right\}  
+c_{L2}\left\{
\left(\frac{k_{\rm t1}}{k_{\rm S}} \right)^2
+\left(\frac{k_{\rm t2}}{k_{\rm S}} \right)^2
\right\}.
\label{bdl}
\eeq
The continuity of the conduction band at the saddle point is of no importance, 
since only the combined (joint) DOS between the conduction band and the valence one 
with the total momentum zero dominates the degree of the resonance of the 
photo-excited electron-hole pair.

For the valence band, we consider only the heavy hole band, 
since the light hole band is irrelevant for our present purpose.
We parametrize the band dispersion for the $\Gamma$ valley region (
$0 \le k_{\rm L}, k_{\rm t1},k_{\rm t2} \le k_{\rm S}$) as
\beq
E_{\rm h} ({\bm k})=c_h \left\{
\left(\frac{k_{\rm L}}{k_{\rm S}} \right)^2 + 
\left(\frac{k_{\rm t1}}{k_{\rm S}} \right)^2+
\left(\frac{k_{\rm t2}}{k_{\rm S}} \right)^2
\right\},
\label{bdhg}
\eeq
while for the L valley region ($k_{\rm S} \le k_{\rm L} \le \pi$, $0 \le k_{\rm t1},k_{\rm t2} \le k_{\rm S}$) as
\beq
E_{\rm h} ({\bm k})=E_{g3}-d_h \left( \frac{\pi-k_{\rm L}}{\alpha_L \pi}\right)^2 
+c_{h} \left\{
\left(\frac{k_{\rm t1}}{k_{\rm S}} \right)^2
+\left(\frac{k_{\rm t2}}{k_{\rm S}} \right)^2
\right\}.
\label{bdhl}
\eeq

We set
$\alpha_S=0.3059$, $\alpha_L=0.6941$, 
$E_g=1.45$, $E_{g2}=1.75$, $E_{g3}=1.40$,
$c_g=1.1$, $c_{L1}=0.80$, $c_{L2}=1.1$,
$c_h=0.50$, $d_L=2.0$, and $d_h=0.90$. 
The resultant band dispersion along $\Gamma$-L line is 
shown in Fig. 1. 
It should be noted that there exist eight equivalent L points 
in the first Brillouin zone of GaAs, and this fact is taken into account 
in our calculation.

\section{Exciton wave function in the Fourier space and the real space}\label{sec:fourier}

Here, we explain the relation between the Fourier (${\bm k}$) space and the real (or relative) space 
of the photo-excited exciton wave function with no total momentum.

From Eq. (\ref{eigg}), the $\Gamma$ exciton is given as 
\beq
\mid \Gamma \rangle =N^{-\frac{1}{2}} \sum_{\bm k} f_{\Gamma} ({\bm k})
a^{\dagger}_{\bm k} b^{\dagger}_{-{\bm k}} \mid 0 \rangle.
\label{eigg2}
\eeq
By using the lattice-site representations for the above electron and the hole operators,
we can easily rewrite Eq. (\ref{eigg2}) into a following localized excitation form, as
\beq
\mid \Gamma \rangle = N^{-\frac{3}{2}} \sum_{{\bm k}, {\bm l}, {\bm \Delta}{\bm l}}
f_{\Gamma} ({\bm k}) e^{-i{\bm k} \cdot {\bm \Delta}{\bm l}} a^{\dagger}_{\bm l} b^{\dagger}_{{\bm l}+{\bm \Delta}{\bm l}}
\mid 0 \rangle,
\label{fourier}
\eeq
wherein the electron is at a lattice site ${\bm l}$, while the hole is at a site ${\bm l}+{\bm \Delta}{\bm l}$,
keeping ${\bm \Delta}{\bm l}$ distant from the electron.
It is needless to say that this ${\bm \Delta}{\bm l}$ is the relative space of the electron-hole pair.
From this Eq. (\ref{fourier}),
we can formally derive the photo-excited exciton wave function 
($\equiv w({\bm \Delta}{\bm l})$) in the real (or relative) space ${\bm \Delta}{\bm l}$ as
\beq
w({\bm \Delta}{\bm l}) \equiv N^{-1} \sum_{\bm k} f_{\Gamma} ({\bm k}) e^{-i{\bm k} \cdot {\bm \Delta}{\bm l}} .
\label{realwf}
\eeq
Its normalization condition is as follows:
\beq
\sum_{{\bm \Delta}{\bm l}} \mid w({\bm \Delta}{\bm l})\mid^2=1.
\label{normwf}
\eeq
Using this $w({\bm \Delta}{\bm l})$ for Eq. (\ref{fourier}),
we get the ordinary exciton in the real space as
\beq
\mid \Gamma \rangle =
 N^{-\frac{1}{2}} \sum_{{\bm l}, {\bm \Delta}{\bm l}}
w({\bm \Delta}{\bm l})  a^{\dagger}_{\bm l} b^{\dagger}_{{\bm l}+{\bm \Delta}{\bm l}}
\mid 0 \rangle.
\eeq
Thus, we can see the Fourier component $f_{\Gamma} ({\bm k})$ or its abosolute value has a key information
of the exciton. This is common to the other photo-excited exciton wave functions with no total momentum.

\end{document}